# Implementation and optimization of Wavelet modulation in Additive Gaussian channels

R. Niazadeh, S. Nassirpour, M.B. Shamsollahi

Electrical Engineering Department, Sharif University of Technology, Tehran, Iran

*Abstract—* In this paper, we investigate the implementation of wavelet modulation (WM) in a digital communication system and propose novel methods to improve its performance. We will put particular focus on the structure of an optimal detector in AWGN channels and address two main methods for inserting the samples of the message signal in different frequency layers. Finally, computer based algorithms are described in order to implement and optimize receivers and transmitters.

*Keywords* Wavelet Modulation;

## I. Introduction

Various analytical methods have been suggested for joint time-frequency analysis. Among these analytical tools, the wavelet transform is of particular interest due to its distinctive properties that make it usable not only in signal processing applications, but also in telecommunication applications; for example for sending signals in a fractal modulation framework, in which we can send several versions of the message in different frequency layers and therefore transmit the data with a lower bit error rate. This modulation can guarantee a reliable communication through channels, in which we do not have any information about their exact band-width, bit interval and frequency properties. This powerful method has the property that if the message signal gets distorted in the channel due to different undesirable phenomena like ISI and fading, the information can still be retrieved in the receiver by demodulating the redundant data existing in other rates. In addition to all these, due to its fractal characteristics, a wavelet modulated signal is noise like and hence can be used in secure data transmission. The Wavelet Modulation (WM) can supersede error control coding methods in wireless communication systems to eliminate undesirable effects resulted from various phenomena including Doppler's effect, multipath effect, etc. [1].

## II. Implementation

In Wavelet Modulation we construct a self-similar signal by adding together a countable number of scaled and modulated versions of our message signal. The discrete wavelet series of a signal is the key tool for implementing the Wavelet Modulation which is defined as follows [2]:

$$s(t) = \sum_n c_{j_l,n}\, \varphi_{j,n}(t) + \sum_{j=j_l}^{j_u} \sum_n x_j(n)\, \psi_{j,n}(t) \quad (1)$$

$$c_j(n) = <\varphi_{j,n}(t), s(t)>, \varphi_{j,n}(t) = 2^j \varphi(2^j t - n) \quad (2)$$

$$x_j(n) = <\psi_{j,n}(t), s(t)>, \psi_{j,n}(t) = 2^j \psi(2^j t - n) \quad (3)$$

Where $\varphi(t)$ and $\psi(t)$ are the scaling and wavelet functions respectively [3], and $\{x_j(n)\}$ are different versions of the message signal inserted in the modulated signal, $s(t)$. It can be shown that if a signal in $L^2(R)$ is homogenous (which means that its scaled versions are proportional to its original version), its detail coefficients would be the same in all different scales [2]. Therefore, if $V_j$ is supposed to be the set of MRA subspaces with $W_j$ as the difference between these subspaces, and $c_{j_l}s$ are arbitrarily selected to be zero, then $s(t)$ can be considered as a signal in $V_{j_u+1}$ subspace with zero projection on $V_{j_l}$. This modulated wave is transmitted through the communication channel. The redundancy of data in this signal allows the recovering of the message signal in receiver with less bit error rate.

A fast algorithm for wavelet decomposition of the modulated signal at the receiver is the Mallat algorithm [4] which can be implemented using a filter bank structure, as described below:

$$c_j(n) = \sum_m h(m-2n)\, c_{j+1(m)} \quad (4)$$

$$x_j(n) = \sum_m g(m-2n)\, c_{j+1(m)} \quad (5)$$

Where $\{c_j(n)\}$ is the projection of the received signal on $\{V_j(n)\}$ and h(n) and g(n) are the scaling and quadrature filters, respectively. Figures 1 and 2 present this implementation (the coefficients $C_{J_{u+1}}(n)$ are the projections of the modulated signal on $V_{J_u+1}$ which, due to the similarity of $\varphi(2^{J_u+1}t)$ and the Dirac delta function, are approximately identical to the samples of the received signal when sampled at a rate of $R = 2^{J_{u+1}}$):

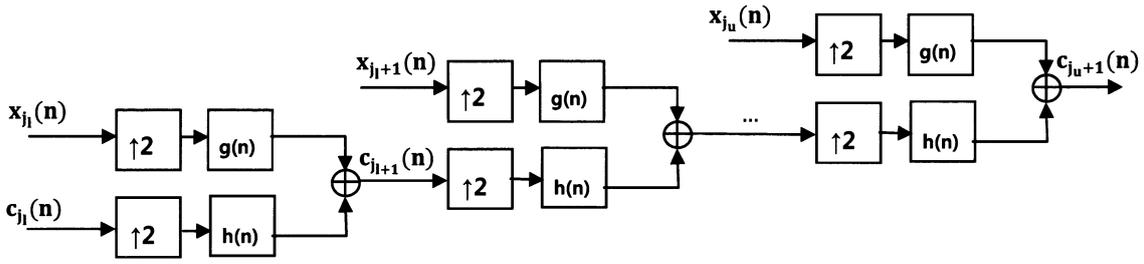

Fig.1: Wavelet modulator

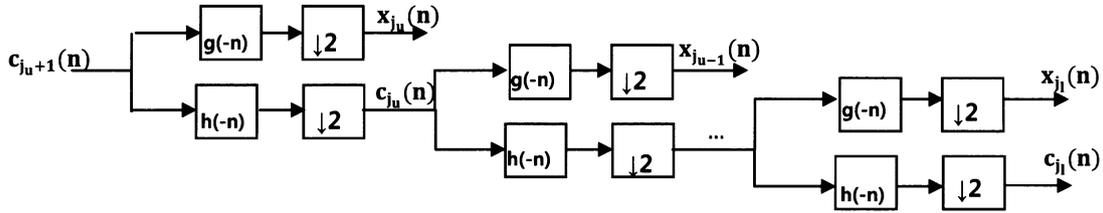

Fig.2: Wavelet demodulator

Since there are decimation blocks in this structure, it is more efficient to put a version of the message signal with fewer samples in lower scales to reduce the amount of memory required to implement the structure. In this paper we will address two novel and efficient methods, which to our best knowledge are proposed for the first time. In the first method, we will put decimated versions of the $2^L$-point signal $x(n)$ in different scales (i.e. $x_{j(n)} = x(2^{j_u-j}n)$, where $x(n)$ is the message signal). In the second method, we will put the L-point signal $x(n)$ with $2^m$ time repetitions in $x_{j_l+m}(n)$ (i.e. $x_{j(n)} = x(n\ mod\ l), n \in \{0,1,2,...,2^{j-j_l}L - 1\}$ ). In order to send a block of symbols with identical size, less memory is needed in the first method than the second one (L compared to $2^L$), however, in the second method, every transmitted symbol will repeat at all rates.

In order to prove the efficiency of these methods, their performance is examined in an AWGN channel which can be modeled as a limited bandwidth, limited duration channel. We have considered a binary communication case in order to simulate the results. i.e. our message signal is a random sequence of binary values and each bit has an energy value of $E_0$, in other words $x(n) \in \{\sqrt{E_0}, -\sqrt{E_0}\}$. $s(t)$ and $r(t)$ are the modulated and the received signal respectively. In an AWGN channel, $r(t) = s(t) + z(t)$, where $z(t)$ is a Gaussian random process with zero mean and a variance of $\delta_z^2$. In the receiver, the projection of $r(t)$ on $V_{j_u+1}$, would build the $r_{j(n)}$ coefficients. The observation vector in the receiver would be as follows:

$$\vec{r}(n) = \{r_{j(n)}: j \in J, n \in N(j)\} \quad , r_{j(n)} = x_{j(n)} + z_{j(n)} \quad (6)$$

$\vec{r}(n)$ consists of the related observation for a specific value of $n$ in different scales. It can be shown that by wavelet-based model for $1/f$ processes (such as $z(t)$ in practice) $z_{j(n)}$s may be modeled as independent Gaussian random variables with zero mean and a variance of $\delta_z^2$[5]. The observation vector would have a different structure according to the algorithm used in the receiver. In the first method, if we assume the block to be sufficiently large, it can be seen that for a specific value of n, the symbols of $x(n)$ would repeat in all scales (in fact, this approximation indicates the upper limit of the probability of error). Therefore, the observation vector would be:

$$\vec{r}(n) = \{r_{j(n)}: j \in J = \{j_l, j_l + 1, ... j_u\}\} \quad (7)$$

It can be proved that the optimum decision rule resulting from the ML would be as followed (assuming that $H_0$ and $H_1$ are equally probable):

$$P(H_0|\vec{r}(n)) \lessgtr_{H_0}^{H_1} P(H_1|\vec{r}(n))$$
$$\Rightarrow l = \sum_{j \in J} \frac{\sqrt{E_0}r_{j(n)}}{\delta_z^2} \lessgtr_{H_0}^{H_1} 0 \quad (8)$$

In the second method, the observation vector can be expressed as:

$$\vec{r}(n) = \{r_{j(n+mL)}: j \in J = \{j_l, j_l + 1, ... j_u\}, m \in \{0,1,2, ... 2^{j-j_l} - 1\}\}$$
$$n \in N(j) = \{0,1,2, ..., 2^{j-j_l}L - 1\} \quad (9)$$

And the optimum decision rule would be modified as follows:

$$P(H_0|\vec{r}(n)) \lessgtr_{H_0}^{H_1} P(H_1|\vec{r}(n)) \Rightarrow$$
$$l = \sum_{j=j_l}^{j_u} \sum_{m=0}^{2^{j-j_l}} \frac{\sqrt{E_0}r_{j(n+mL)}}{\delta_z^2} \lessgtr_{H_0}^{H_1} 0 \quad (10)$$

With this decision rule, the more the repetitions of a symbol are, the less the probability of error is. So, in sending blocks of the same length, the second method would have lower error probability but also With this decision rule, the more the repetitions of a symbol are, the less the probability of error is. So, in sending blocks of the same length, the second method would have lower error probability but also lower spectral efficiency. When using the first method, the probability of error can be formulated as below:

$$P_\varepsilon = P(I > 0|H_1) = Q\left(\sqrt{\frac{ME_0}{\delta_z^2}}\right) \quad (11)$$

$$M = j_u - j_l + 1, \quad Q(x) = \int_x^\infty \frac{1}{\sqrt{(2\pi)}} e^{-x^2/2} dx$$

And when using the second method, it would be:

$$P_\varepsilon = P(I > 0|H_1) = Q\left(\sqrt{\frac{KE_0}{\delta_z^2}}\right) \quad (12)$$

$$K = \sum_{j=j_l}^{j_u} 2^{j-j_l}, \quad Q(x) = \int_x^\infty \frac{1}{\sqrt{(2\pi)}} e^{-\frac{x^2}{2}} dx$$

The drawback of the wavelet modulation compared to other digital modulations is its relatively low spectral efficiency ($\eta_f = 0.5$) [2]. In figure 3, the probability of error vs. SNR is plotted for the first and the second method together with the PAM modulation for comparison. The relative data is resulted from the simulation of a 512-point signal transmission in 6 successive scales and the wavelet used in this simulation is Daubechies (N=4). As these figures imply, the first and the second method both have a significant SNR improvement compared to PAM. But, the amount of memory required to implement the first method is remarkably less than the second one. These results are summarized in Table 1.

Table I   Comparison of the Two Methods

| Method | Simulation Time [1] | SNR improvement in comparison with PAM at probability of error = 0.1 |
|---|---|---|
| WM1 | 2.3s | 18.7dB |
| WM2 | 50.22s | 13.4dB |

[1]*Indicates the amount of memory required to implement the method*

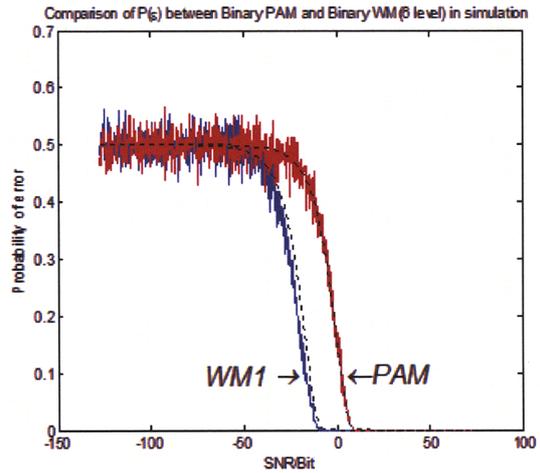

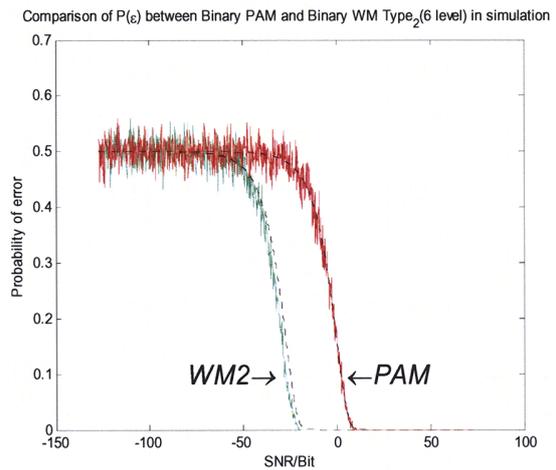

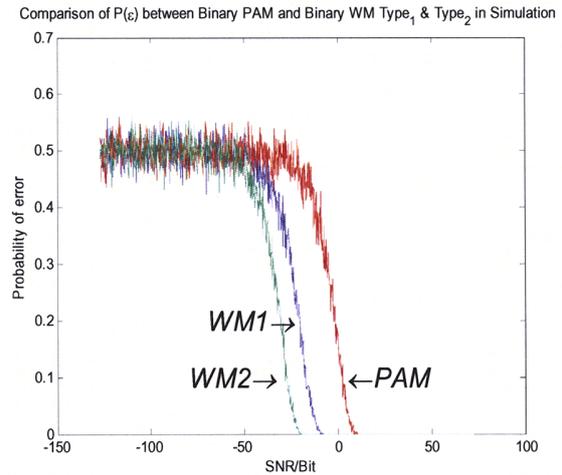

Figure1

References:

[1] Manglani, M. J. and Bell, A. E., "Wavelet modulation in Gaussian and Rayleigh fading channels", Acoustics, Speech, and Signal Procesing Proceeding(ICASSP'02), IEEE International Conference, 2002

[2] Oppenheim, A. V. and Wornell, G. W., "Wavelet-based representation for a class of self-similar signals with application to fractal modulation", IEEE Transactions on Information Theory, 38(2):785-800, 1992

[3] Burrus, C. S., "Introduction to wavelets and wavelet transforms", New Jersey: Prentice Hall, 1998

[4] Mallet, S. G., "A theory for multiresolution signal decomposition: The wavelet representation", IEEE Transactions on Pattern Analysis and Machine Intelligence, 2(7):674-692, 1989

[5] Wornell, G. W., "A Karhunen-Loeve like expansions for 1/f processes via Wavelets", IEEE Transactions on Information Theory, 33: 859-861, 1990